\documentclass[journal]{IEEEtran}

\usepackage{bm}
\usepackage{graphicx}
\usepackage{epstopdf}
\usepackage{amsmath, amsthm, amsfonts, amssymb, amsbsy}
\usepackage{setspace}
\usepackage{pstricks,arydshln}
\usepackage{multirow}
\usepackage{booktabs}
\usepackage{stfloats}
\usepackage{amscd,textcomp,amssymb,epsfig,graphics,epsf,color,amsmath,balance,cite}
\usepackage{subfigure}
\usepackage{caption}
\usepackage{cite}
\usepackage{xcolor}
\usepackage{epstopdf}
\usepackage{cases}
\usepackage{pmat}
\usepackage[noend]{algpseudocode}
\usepackage{algorithmicx,algorithm}

\newcommand{\RNum}[1]{\uppercase\expandafter{\romannumeral #1\relax}}
\usepackage{threeparttable}
\usepackage{booktabs,arydshln, multirow}

\begin{document}

\title{Smart Roads: Roadside Perception, Vehicle-Road Cooperation and Business Model}

\author{Rui Chen,~\IEEEmembership{Member,~IEEE,} Lu Gao,~\IEEEmembership{Graduate Student Member,~IEEE,} Yutian Liu,~\IEEEmembership{Graduate Student Member,~IEEE,} Yong Liang Guan,~\IEEEmembership{Senior Member,~IEEE,} and Yan Zhang,~\IEEEmembership{Fellow,~IEEE}

\thanks{Rui Chen is with the State Key Laboratory of ISN, Xidian University, Xi'an 710071,
China, and also with the Guangzhou Institute of Technology, Xidian University, Guangzhou 510700, China (e-mail: rchen@xidian.edu.cn).}
\thanks{Lu Gao is with the Guangzhou Institute of Technology, Xidian University, Guangzhou 510700, China (e-mail: lugao0921@stu.xidian.edu.cn).}
\thanks{Yutian Liu is with Urbanning Planning and Transportation, Eindhoven University of Technology, 5600 MB, Netherlands (e-mail: yutian\_liu@hotmail.com).}
\thanks{Yong Liang Guan is with the School of Electrical and Electronic Engineering, Nanyang Technological University, 639798, Singapore (e-mail: EYLGuan@ntu.edu.sg).}
\thanks{Yan Zhang is with the Department of Informatics, University of Oslo, 0316 Oslo, Norway, and also with the Simula Metropolitan Center for Digital Engineering, 0167 Oslo, Norway (e-mail: yanzhang@ifi.uio.no).}
}

\maketitle

\begin{abstract}

Smart roads have become an essential component of intelligent transportation systems (ITS). The roadside perception technology, a critical aspect of smart roads, utilizes various sensors, roadside units (RSUs), and edge computing devices to gather real-time traffic data for vehicle-road cooperation. However, the full potential of smart roads in improving the safety and efficiency of autonomous vehicles only can be realized through the mass deployment of roadside perception and communication devices. On the one hand, roadside devices require significant investment but can only achieve monitoring function currently, resulting in no profitability for investors. On the other hand, drivers lack trust in the safety of autonomous driving technology, making it difficult to promote large-scale commercial applications. To deal with the dilemma of mass deployment, we propose a novel smart-road vehicle-guiding architecture for vehicle-road cooperative autonomous driving,  based on which we then propose the corresponding business model and analyze its benefits from both operator and driver perspectives. The numerical simulations validate that our proposed smart road solution can enhance driving safety and traffic efficiency. Moreover, we utilize the cost-benefit analysis (CBA) model to assess the economic advantages of the proposed business model which indicates that the smart highway that can provide vehicle-guided-driving services for autonomous vehicles yields more profit than the regular highway.

\end{abstract}

\section{Introduction}
\begin{figure}[t]
\begin{center}
\includegraphics[width=8.6cm]{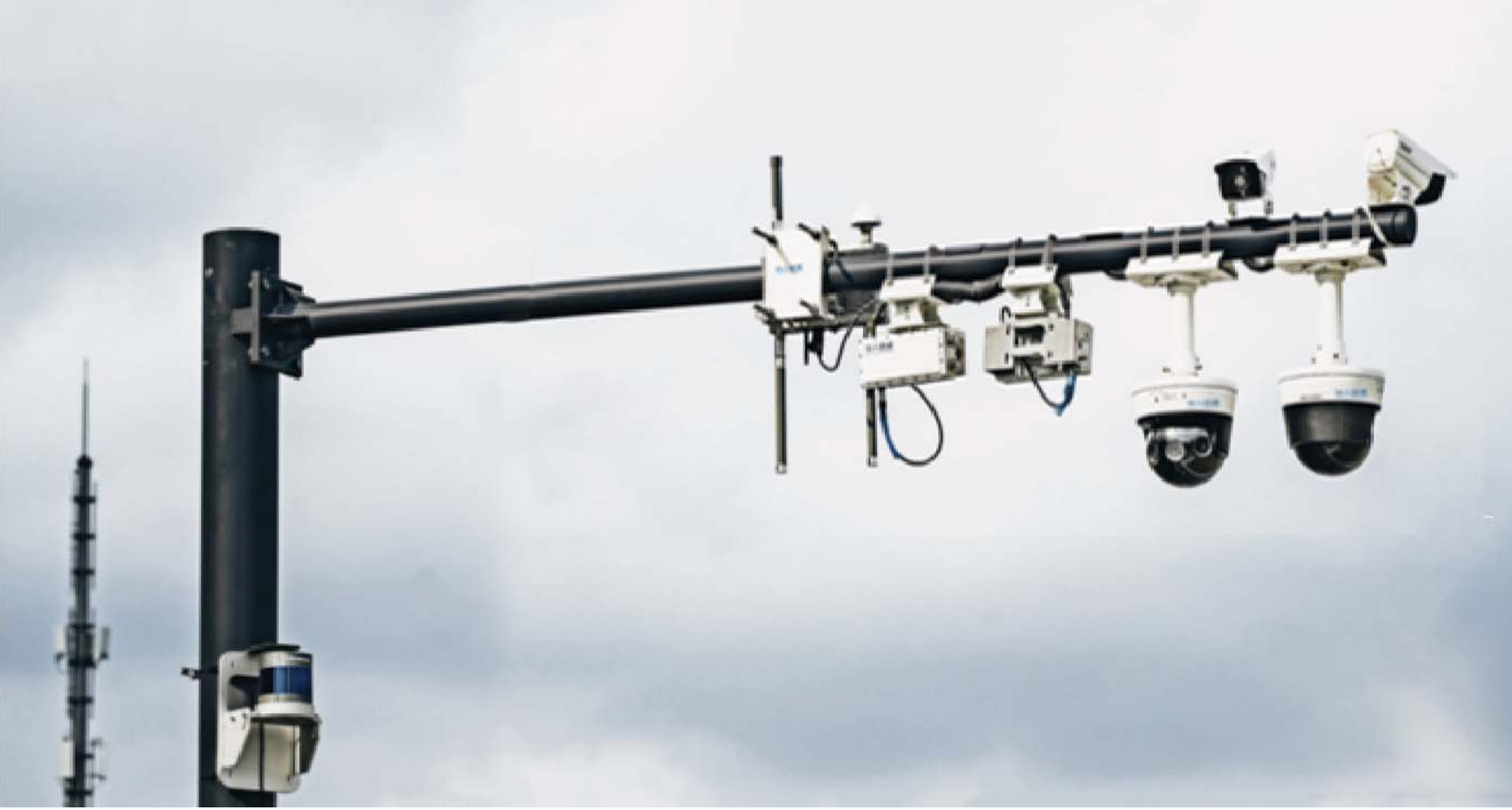}
\end{center}
\caption{A practical deployment of roadside perception sensors including cameras, lidar, mmWave radars and RSUs.}%
\label{fig1}
\end{figure}

\IEEEPARstart{W}ITH  the rapid advancement of Internet of things (IoT) and artificial intelligence (AI) technologies, smart roads have emerged as an indispensable component of intelligent transportation systems (ITS). Smart roads can be defined as road infrastructure that is integrated with advanced network and communication technologies. As shown in Figure \ref{fig1}, roadside perception utilizes various sensors such as visual sensors, mmWave radar, and LiDAR, combined with edge computing devices, to acquire real-time information on current traffic data. Research on smart roads, specifically automated highway systems, was initiated by the United States in the 1980s, with the Federal Highway Administration taking a leading role in these exploratory efforts \cite{bender1991overview}. In Europe, the Easyway project was launched to cover 30,000 kilometers of highways with cross-border data exchange \cite{rafiq2013s}. In Japan, the ``Smartway'' project was proposed in 2004, building upon the vehicle infrastructure communication systems  (VICS) project \cite{arino2008its}. 

Especially in China, the government has included the development of smart road infrastructure in its national development plan and allocated significant funds to support the implementation. For example, in the expansion project of the highway between Changsha and Yiyang, a continuous arrangement of cameras and millimeter-wave radars dynamically monitors road information, while the data is processed through edge computing \cite{zhao2021design}. The Beijing section of the YanChong Expressway has emerged as a demonstration route for the Winter Olympics' transportation, with a Z-shaped deployment of C-V2X RSUs, cameras, millimeter-wave radars, etc., enabling real-time dialogue between vehicles and roadside multi-source data to support autonomous driving \cite{liu2023research}. This facilitates seamless connectivity among roadside perception devices, vehicle terminals, and cloud platforms, offering comprehensive concierge services to drivers.

Unfortunately, roadside perception data is underutilized, limiting autonomous vehicles' adaptability and safety. To unlock smart road benefits, widespread deployment of perception devices is crucial despite initial investment challenges and driver trust issues. This paper proposes a smart-road vehicle-guiding architecture, enhancing safety, efficiency, and cost-effectiveness. A corresponding business model is explored through quantitative analysis, benefiting all stakeholders. Simulation experiments validate the feasibility, offering valuable insights for policymakers, industry stakeholders, and researchers in autonomous driving.

The remainder of the paper is organized as follows.
Section \RNum{2} reviews the roadside perception sensors and data fusion.
Section \RNum{3} analyzes the mass deployment dilemma in detail.
Section \RNum{4} proposes a smart-road vehicle-guiding architecture and a new business model.
The numerical simulations and results are reported in Section \RNum{5}. Finally, Section \RNum{6} concludes the research and presents the future research directions.

\section{Roadside Perception Sensors and Data Fusion}
The realm of smart road is increasingly focusing on roadside perception sensors and data fusion technology. This section delves into these two pivotal technologies from various perspectives. The advancements in these cutting-edge technologies also furnish us with a robust technological underpinning to construct smart-road vehicle-guiding architecture for vehicle-road cooperative autonomous driving.
\subsection{Roadside Perception Sensors}
The advantage of roadside perception sensors is the ability to expand the perception range of autonomous vehicles. Currently, roadside perception mainly relies on AI-driven visual camera and mmWave radar solutions. LiDAR and radar video integration is a growing trend in roadside perception.
\subsubsection{AI-Driven Visual Camera}
The integration of roadside cameras with the visual artificial intelligence analysis function has enabled a more intuitive display of the traffic status. Currently, vendors such as Huawei, Dahua Technology, and Hikvision have rolled out their AI-driven roadside cameras.

Huawei's AI ultra-micro light camera combats light pollution and image quality issues. Employing DNN ISP image optimization, it captures clear images of license plates, vehicle colors, and passengers even in low light. Its open architecture SDC OS enables swift integration of third-party algorithms via the Huawei HoloSens Store, offering versatility across multiple scenarios with robust computing power.
\subsubsection{MmWave Radar}
MmWave Radar boasts numerous advantages. It excels in accuracy, wide detection range, and penetration, unaffected by adverse weather. This radar system tracks multiple targets continuously and precisely, utilizing mmWave radar beamforming to identify and monitor vehicles and pedestrians, providing crucial data on distance, speed, angle, and direction. Multiple vendors invest in improving mmWave radar performance.

Hurys unveils a new-gen wide-area mmWave radar perception system with advanced front-end and data processing technologies, offering richer data \cite{wang2023realtime}. In 2020, Muniu Technology launched WAYV ultra-long-range mmWave radar, setting a market record with a 1,000-meter detection range. 

\subsubsection{LiDAR}
LiDAR offers precise 3D data, allowing e-fence control, target filtering, and tailored communication in specific areas. Integrating LiDAR into roadside perception enhances traffic monitoring's accuracy and efficiency. 

Traditional roadside solution providers like VanJee Technology, Changsha Intelligent Driving Institute (CiDi), and China TransInfo Technology offer their LiDAR products. In March 2021, VanJee Technology deployed smart base stations with V2X roadside antennas and LiDAR in Xiongan Civic Center V2X and High-speed Railway Hub Road Intelligence Projects.

\subsubsection{Radar Video All-in-One}
Integrating radar and video tech into unified systems cuts material and installation costs. This combo boosts target detection precision and deploys fusion algorithms to reduce latency and edge-end computing load.

Currently, vendors specializing in roadside vision-based HD such as Dahua Technology and Hikvision, alongside roadside radar vendors including Raysun Radar, Hurys, and DeGuRoon, have introduced their radar video all-in-one solutions. Oculii's 2019 4D radar video all-in-one, using Falcon, marked a breakthrough. In H2 2021, Raysun Radar launched IET6LRR, a new-gen radar video all-in-one, with an impressive 425-meter detection range.

\subsection{Data Fusion-Based Holographic Perception}
Camera, mmWave radar, and LiDAR possess distinct advantages and disadvantages. Therefore, the integration of multiple sensors is the primary trend in the development of holographic roadside perception \cite{xu2019review}. Since 2000, several companies, including Huawei, Baidu, OriginalTek, and the institute of deep perception technology (IDPT), have introduced their holographic perception solutions.

\subsubsection{Huawei}
In 2020, Huawei unveiled the Holographic Perception Intersection 1.0, followed by Holographic Perception Intersection 2.0 in March 2021. This advanced solution integrates AI ultra-low light cameras, mmWave radar, ITS800 edge computing nodes, and intersection HD maps, suitable for diverse intersections. Huawei's holographic perception system blends, accesses, analyzes, and communicates at intersections through edge computing units, facilitating traffic signal and onboard unit (OBU) integration, laying the foundation for future cooperative vehicle infrastructure system (CVIS) services.

\subsubsection{Baidu}
In March 2021, Baidu unveiled the ACE Intelligent Intersection Solution, enabling the perception of road elements, vehicles, pedestrians, environments, and incidents. It employs cameras, fisheye cameras, LiDAR, and roadside edge computing units. Integrated with Baidu Map data, it achieves exceptional data detection accuracy exceeding 97\%. Collaborating with ecosystem partners, Baidu customizes cameras, LiDAR, communication gear, and edge computing units for the ACE Intelligent Intersection Solution.

\subsubsection{IDPT}
The holographic traffic perception solution proposed by IDPT involves the deep integration of camera, lidar, mmWave radar, and radar video all-in-one. This innovative solution enables front-end multi-device stitching and achieves comprehensive 360-degree image-level holographic perception at all times and in all weather conditions. It provides reliable data for crossroads, with a maximum detection range of up to 500 meters.

\section{The Dilemma of Mass Deployment}
Roadside sensor development is promising, and technologies such as vehicle-road cooperation and data fusion are evolving. However, the current large-scale deployment of roadside perception and communication devices still faces various issues. In this regard, we present a comprehensive analysis that encompasses two perspectives: the funding dilemma and the driver experience dilemma. 

\subsection{Funding Dilemma}
Sustained development of mass deployment of vehicle-road cooperation necessitates substantial funding investment. Firstly, from the perspective of funding sources, the construction and operation of the vehicle-road cooperation require funding support from multiple channels, such as the government, vehicle manufacturers, and communication operators. However, it may be difficult to guarantee the continuous investment of this funding in mass deployment. Government funding may be affected by budget constraints and policy changes, while enterprise funding may be constrained by commercial interests and market competition. 

The cost issue is also a crucial consideration that cannot be overlooked. On one hand, the initial construction costs are immense. The prerequisite for the effectiveness of the vehicle-road cooperation system lies in the sufficient deployment of infrastructure at the roadside, including roadside sensors, communication devices, and edge computing servers, as well as the development and testing of related software. On the other hand, the long-term maintenance and operational costs also demand consideration. With the increasing number of vehicles, the system's capacity and coverage need to be expanded accordingly. The related infrastructure requires regular upgrades and maintenance. Moreover, its operational cost includes data storage, processing, transmission, and maintenance, all of which require continuous investment.

Finally, we should consider the issue of return on investment when it comes to massive capital injection. Although the current vehicle-road cooperation can bring benefits in terms of improving traffic efficiency, reducing carbon emissions, and enhancing traffic safety, the return on these benefits may take a long time to materialize, posing a challenge to the return on investment. In addition, there is no direct business model proposed to enable related operators to directly benefit from the system.

\subsection{Driver Experience Dilemma}
Driver experience is pivotal for large-scale vehicle-road cooperation success. As drivers serve as direct contributors to economic benefits, it is necessary to provide more intelligent, personalized, and comfortable travel experiences. Undeniably, the vehicle-road cooperation system can improve road traffic and reduce accidents. However, from the perspective of drivers, their sense of experience is not very adequate. The main issues manifest in limited autonomy, technological constraints, poor interaction experience, and system malfunctions and unreliability. These problems restrict the vehicle-road cooperation system from achieving complete liberation from human driving constraints, necessitating technological improvements and innovations to enhance the quality of the drivers' experience. 

In addition, the success of vehicle-road cooperation relies on the collection and analysis of vast amounts of traffic data, which may compromise users' privacy. Without stringent data protection measures, users may be concerned about their data being leaked or misused, and therefore hesitant to adopt vehicle-road cooperation. Thus, ensuring data security and guaranteeing a positive user experience must be considered for the success of vehicle-road cooperation. The high-quality driver experience can create a virtuous cycle, benefiting both drivers and operators and leading to continued customer acquisition and service improvements.

\section{Solutions}
Based on the analysis of these dilemmas, we propose a novel smart-road vehicle-guiding architecture for autonomous vehicles and put forward an impending business model to enable practical application and commercialization.

\subsection{The Smart-Road Vehicle-Guidance Architecture}
\begin{figure}[t]
\begin{center}
\includegraphics[width=8.6cm]{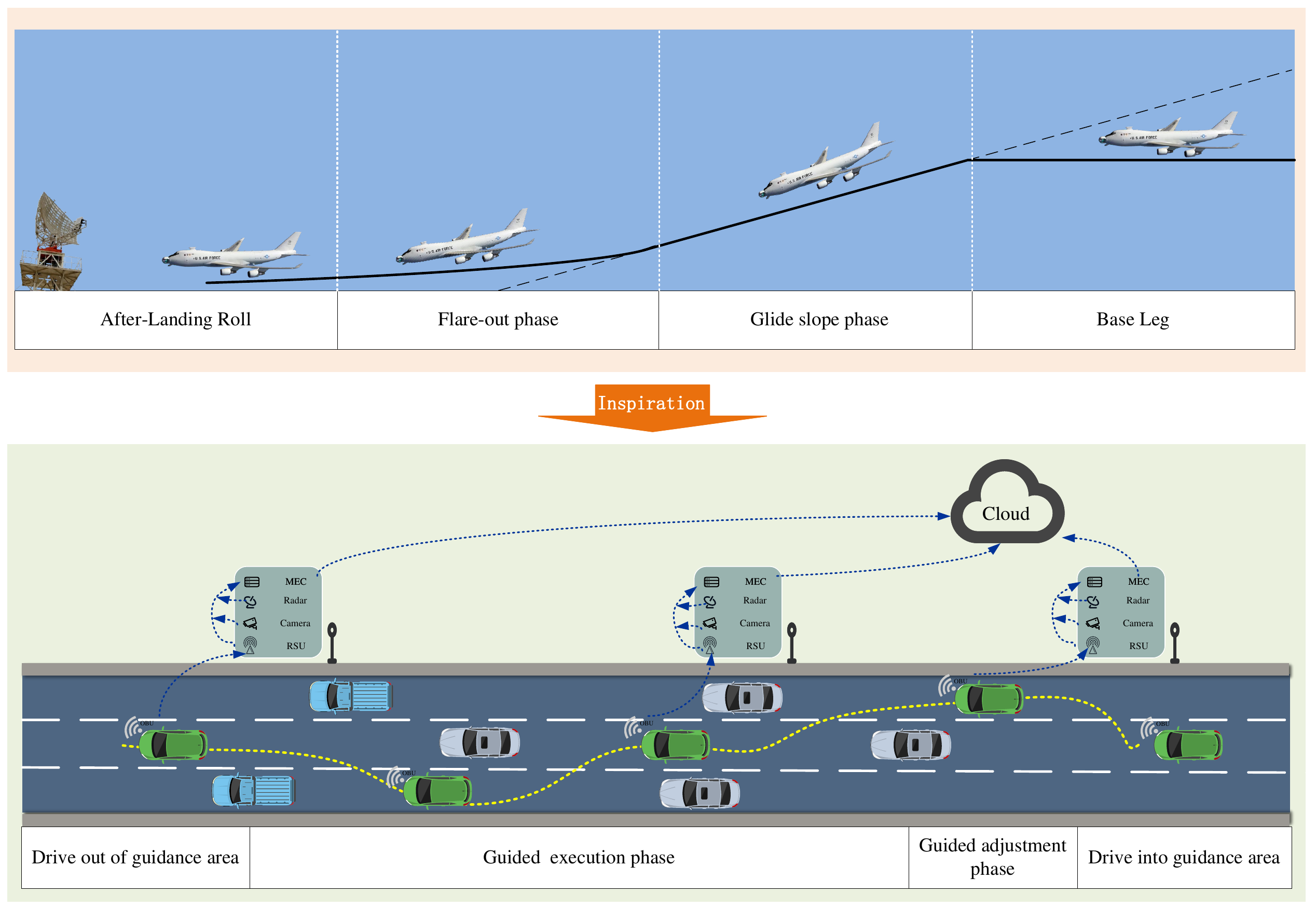}
\end{center}
\caption{Schematic diagram of guided vehicle autonomous driving based on smart road.}%
\label{fig3}
\end{figure}

The design of our proposed architecture is inspired by aircraft landing radar guidance systems. As shown in Figure \ref{fig3}, taking into account the technical characteristics and processes of the aforementioned aircraft landing radar guidance system, we design a smart-road vehicle-guiding architecture for autonomous driving. Once a compliant vehicle enters the designated area, it has the option to procure vehicle-guided-driving services for autonomous vehicles offered by the smart road. After opting for the guidance driving services, the vehicle's perception and decision-making primarily rely on the data provided by the roadside sensing devices, supplemented by the data obtained from the vehicle-side perception devices. Throughout the guidance process, cameras, mmWave radars, and meteorological sensors capture and collect video, radar, and meteorological data, respectively. These diverse data inputs are deeply integrated and processed by the edge computing server, offering dependable support for decision-making, control, and interaction within the guidance architecture. Additionally, wireless communication technologies facilitate V2X communication. Before the target vehicle exits the guidance area, the onboard system issues a warning to remind the driver to take control of the vehicle. If the driver takes over and operates the vehicle safely, the vehicle can continue its journey after leaving the guidance area. However, if the driver does not take over the vehicle after a while, the architecture guides the vehicle to temporarily stop at a roadside parking area until the driver regains control of the vehicle before leaving the guidance area.

To ensure the continuity of guidance and provide accurately continuous roadside perception data to guided vehicles, we have employed a segmented and continuous deployment strategy for installing roadside devices. The roadside device deployment scheme employed in this study primarily comprises RSUs (Roadside Perception Units), MECs (Mobile Edge Computing Units), and roadside perception sensors. These devices serve distinct functions for communication, computation, and perception. Roadside perception sensors encompass cameras and millimeter-wave radars, each characterized by differing design principles and operational mechanisms, tailored to varying scenarios. For dual 6-lane highways, we symmetrically position deployment points on both sides, with an interval of 400 meters between each point. Along each kilometer stretch, 2.5 points are unidirectionally deployed, while 5 points are bidirectionally deployed. The horizontal deployment of roadside devices is situated directly above the lanes of unidirectional traffic.

\subsection{The Proposed Business Model}
Based on the smart-road vehicle-guiding architecture for autonomous driving, this paper proposes a novel business model. Specifically, our business model relies on charging guided fees during the process of smart road guiding smart vehicles for autonomous driving to achieve profitability. To validate the feasibility of this new business model, we analyzed from two perspectives: operators benefits and drivers benefits.

\subsubsection{Operators Benefits}
From operators perspective, embracing a novel business model necessitates contemplating its economic value. Realizing substantial economic value entails two key aspects: cost reduction and revenue increase. Concerning revenue increase, this model surpasses traditional roads by incorporating guided fees, thereby yielding a considerable supplementary income. Furthermore, this novel business model has the capability to effectively reduce costs. Undeniably, the installation, deployment, and maintenance of roadside perception and communication devices entail substantial expenses. Fortunately, with the advancement of smart transportation, vehicle-road cooperation devices evolve from the laboratory stage to the functional mass production stage. The cost of infrastructure construction is greatly reduced. Moreover, due to the benefit of industrial upgrading and iteration, the industry cost also continues to decrease. Compared with other investments, the service life of road infrastructure is quite long if the firmware and software are upgraded periodically. 

Next, we analyze the cost of the smart guided vehicles used in this model. With the high-precision traffic environment perception provided by roadside sensors, the installation cost of the smart guided vehicles' perception sensor is greatly reduced.  Traditional "driverless vehicles" require expensive components like Lidar, high-precision GPS, and inertial navigation systems, totaling around \$250,000—10-20 times more than regular vehicles. However, smart guided vehicles do not need to use such expensive sensors. Even upgrading and installing sensors directly on existing vehicles is possible, which has the potential to reduce the cost of vehicles. It reduces the threshold for users to purchase them, thereby increasing the possibility of purchasing services.
\begin{figure}[t]
\begin{center}
\includegraphics[width=8.6cm]{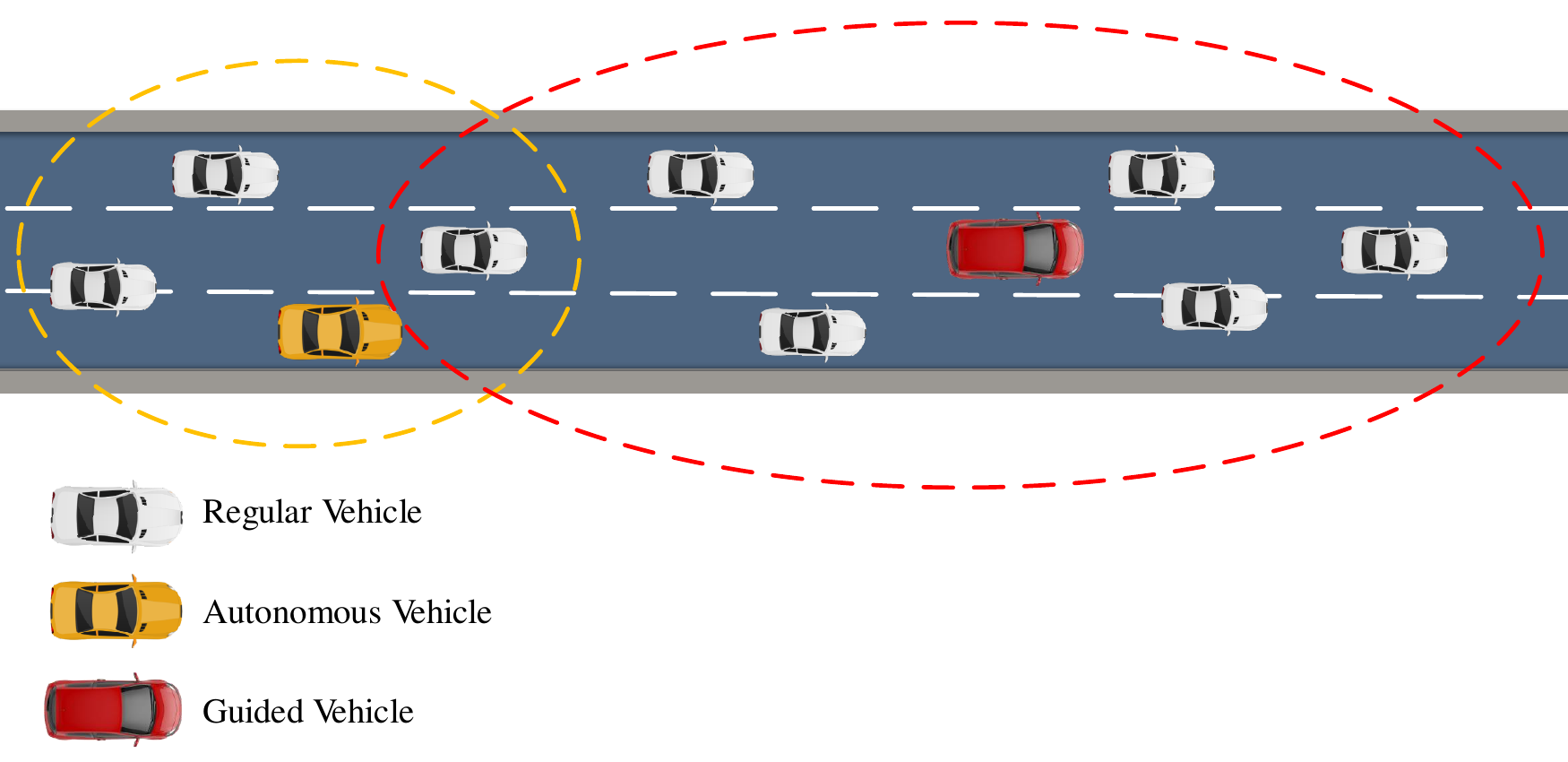}
\end{center}
\caption{Perception range of AV and GV.}%
\label{fig4}
\end{figure}
\subsubsection{Drivers Benefits}
Prioritizing driver benefits is crucial for service consumers. This model transforms the driving experience by freeing drivers from driving tasks, allowing them to relax, be entertained, or work. Using real-time road conditions, the system plans optimal routes for faster and safer travel, enhancing the overall driving experience and reducing fuel consumption, vehicle wear, and driving costs.

Undoubtedly, the improvement in safety brought about by this model should not be underestimated. Particularly in sudden and adverse weather conditions such as rain, snow, or fog, the advantages of guided vehicles used in this model become remarkably evident. As far as we know, Tesla's autonomous vehicles encounter challenges in heavy rain or when lane markings are obscured. In contrast, our model leverages the installation position of roadside radars as reference points to dynamically obtain the relative positions of all traffic participants on the current road segment. Utilizing this real-time data and combining it with stored scene images captured in clear weather conditions, stored in a multi-access edge computing (MEC) system, we can construct models and plan safe and reliable driving routes, guiding vehicles to travel securely.

Lastly, this innovative business model enhances transportation efficiency and mitigates the inconvenience caused to drivers by reducing traffic congestion. As the responsibility of driving control transitions from individual drivers to roadside perception and communication devices for guidance and control, notable benefits arise. The model-guided control effectively mitigates the occurrence of phantom traffic jams triggered by human driving errors. It can also improve traffic efficiency by coordinating vehicle paths to form vehicle formations.

\section{Smart Roads Guiding Smart Vehicles: A Win-Win Case Study}
To further validate the benefits of smart-road vehicle-guiding architecture for autonomous driving, a series of simulation experiments are conducted. Firstly, we construct a realistic traffic simulation model based on the vehicle-road cooperative framework, including road networks, vehicles, and roadside perception devices. Then, we compare the safety of guided vehicles (GV) and autonomous vehicles (AV), using the time-to-collision (TTC) as a quantitative indicator. We also compare the traffic efficiency under different GV penetration, using the average speed of the vehicle as a quantitative indicator. Lastly, we utilize the cost-benefit analysis (CBA) model to evaluate the economic benefits of the new business model, thereby validating its economic feasibility.

\subsection{Simulation Design}
\subsubsection{Driving model}
In this experiment, we employ a vehicle model based on the intelligent driver model (IDM) \cite{treiber2000congested}. The IDM is rooted in microscopic traffic flow models, which offer highly detailed vehicle movement. Each vehicle operates as an independent agent following specific control rules based on its environment. This model enables unique control for each vehicle.

Given the substantial influence of lane-changing behavior on traffic safety, capacity, and stability, our simulation encompasses both car-following and explicit lane-changing rules. We incorporate lane-changing rules from the micro-traffic flow-following model \cite{kesting2007general}. Essentially, when a vehicle follows another, it considers lane changes under conditions like slow speed, excessive speed, or close proximity. The chosen target lane accounts for the vehicle's speed and the leading vehicle's state, aiming for smoother and more efficient travel. When opting for a lane change, the vehicle adjusts speed and position for a seamless transition.

\subsubsection{Simulation Design}
This paper employs Matlab to simulate a 1000-meter, bi-directional, two-lane circular road, creating heterogeneous traffic with regular vehicles (RV), autonomous vehicles (AV), and guided vehicles (GV).  In this experiment, RV serve as ordinary traffic participants and provide traffic variables for our testing. Our main comparative objects are AV and GV. Through analyzing the specific scenes of single-car intelligent autonomous driving and smart-road vehicle-guiding architecture for autonomous driving, we find that the biggest difference between the two lies in their perception range. As shown in Figure \ref{fig4}, the perception range of AV is limited to the surrounding vehicles, and there are larger blind spots, making it difficult to evaluate the entire traffic environment. The biggest advantage of GV is the wider perception range. Through roadside sensors, the traffic environment of the current road section can be perceived in a small area, reducing blind spots and ensuring driving safety. 

In our numerical simulation experiment, we store the driving states of each vehicle from the previous time step. AV use neighboring vehicle states for decision-making, while GV employ a wider range to inform decisions based on the entire traffic environment. Furthermore, to better reflect real-world scenarios, we add noise and confidence to the perceptual information. This is because the perceptual accuracy of the vehicle itself has not yet reached 100\% for AV. For GV, on the one hand, the accuracy of roadside sensors is also affected by various environmental factors. On the other hand, the design of the guiding system based on the vehicle-road cooperation framework should also consider communication delays and computing time.

\begin{figure*}[t]
\begin{center}
\includegraphics[width=18cm]{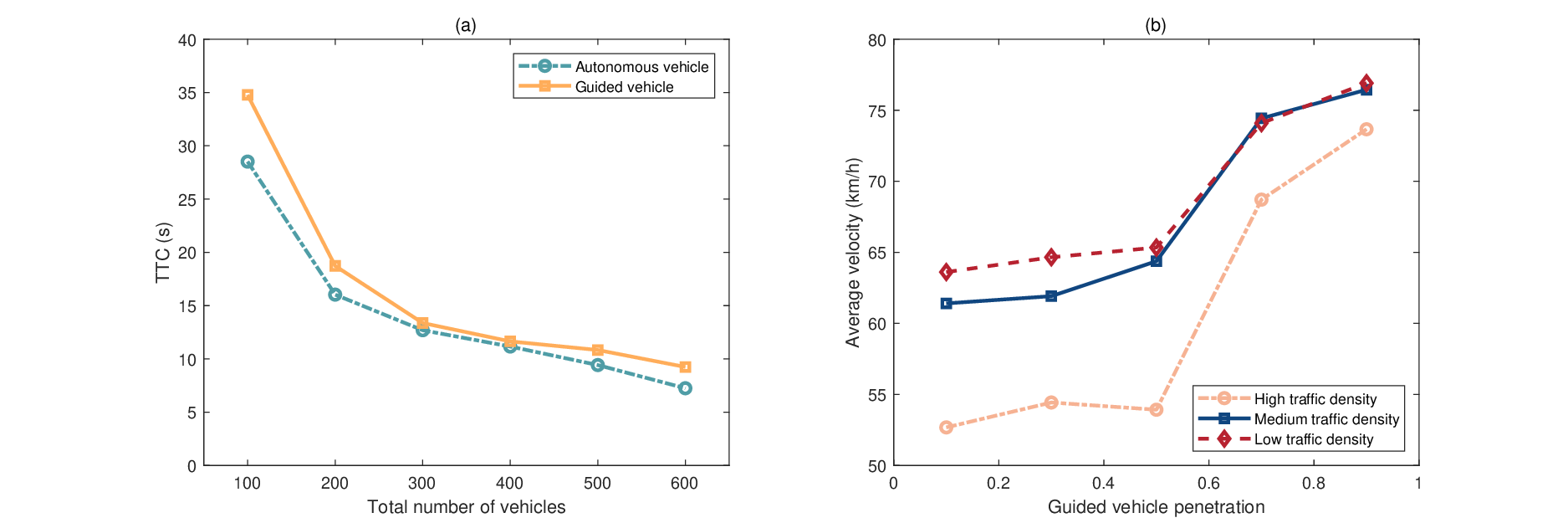}
\end{center}
\caption{Simulation result analysis: a) Comparisons of AVs and GVs TTC  under different traffic environments; b) Comparisons of average velocity under different guided vehicle penetrations in different traffic densities. }%
\label{fig6}
\end{figure*}

\subsection{Simulation Result Analysis}
\subsubsection{Safety}
Safety is paramount in evaluating autonomous driving, and we use Time-to-collision (TTC), a key traffic safety measure, to assess real-time rear-end collision risk \cite{vogel2003comparison}. In our experiment, we employ various models to simulate different vehicle counts, reflecting diverse traffic scenarios. As vehicles increase, the complexity rises, elevating accident risks. We conduct six experiments on the same simulated road with 100, 200, 300, 400, 500, and 600 vehicles, maintaining a 20\% AV, 20\% GV, and 60\% RV ratio for experiment consistency. We calculate mean TTC for AV and GV in each scenario, as depicted in Figure \ref{fig6}a.

As shown in Figure \ref{fig6}a, in different traffic environments, the TTC of GV is higher than that of AV, further indicating that GV has a higher level of safety. In addition, the experimental results show that as the traffic environment becomes more complex (i.e., as the total number of vehicles on the road increased), the TTC of AV and GV decreased. This suggests that as the number of vehicles increases, traffic safety decreases, which is consistent with our predicted results. However, as the traffic environment becomes more complex, the reduction in collision time for GV is greater than that for AV vehicles, indicating that the stability of GV in complex environments still needs improvement.

\subsubsection{Traffic Efficiency}
Various indicators, including total travel time, average waiting count, and average delay time, can assess road segment traffic efficiency \cite{zhu2020safe}. Here, we focus on vehicle average velocity as the key indicator; higher velocity signifies improved traffic efficiency. The evaluation of traffic efficiency using the average velocity of vehicles is depicted in Figure \ref{fig6}b. As the penetration rate of GV increases, the average velocity of vehicles also rises. At high traffic density, the numerical changes are not significant for penetration rates below 0.5, but they rapidly increase beyond 0.5. With the continuous increase in traffic density, the variations in average vehicle velocity become more pronounced, and the impact of GV penetration rate becomes more significant. This indicates that GV have a more significant and positive influence on traffic efficiency, particularly under high traffic density conditions.

\subsection{Economic Benefit Evaluation}
This paper not only conducts a quantitative analysis of the benefits at the technical level, but also evaluates the feasibility of the new business model from an economic perspective. To conduct a comprehensive assessment, we employ the cost-benefit analysis (CBA) methodology to evaluate the cost, benefit, and cost-benefit ratio \cite{mechler2016reviewing}. This paper presents a comparative analysis of the economic benefits between regular highway and smart highway. Smart highway are defined as highway that can provide guidance services for autonomous driving. It is noteworthy that We focus on the Guanghe Expressway Guangzhou section as the research subject, which spans a total length of 70.754 kilometers.

\paragraph{Cost}
When upgrading an existing highway, costs encompass maintenance, operation, and deploying smart roadside perception and communication devices. Maintenance and operation costs, as per the "National Toll Road Statistics Bulletin for 2021" by China's Ministry of Transport, totaled CNY 135,541.61 million for 161,200 kilometers of highways in 2021, averaging about CNY 840,830 per kilometer. We use this as the baseline for maintenance and operation costs. These costs rise by 20\% every three years based on the baseline expenses \cite{anastasopoulos2010cost}.

For smart highways using our proposed business model, we consider higher future expenses for smart roadside devices. Annual maintenance and operation expenses are set 20\% higher than regular highways. Smart roadside perception and communication devices follow a five-year upgrade cycle. Initial additional costs are 10\% of deployment costs, with a 5\% annual increase. Based on data from China CITIC Securities \cite{norden2019efficient}, the 2020 average transformation cost per kilometer was approximately CNY 397,200. Thus, the deployment and installation cost for Guanghe Highway's smart devices amounts to CNY 28.1034888 million.

\paragraph{Benefit}
We begin by examining the relationship between annual traffic flow and revenue data. For the Guanghe Highway Guangzhou section, between 2013 and 2019, the revenue's compound annual growth rate (CAGR) stands at 18.14\%, with a CAGR of 13.43\% from 2017 to 2019, alongside an annual traffic flow growth rate of 11.14\%. As of 2022, this highway records an annual traffic flow of 19.5072 million vehicles, generating CNY 503.4649 million in revenue. Using this data, we forecast annual traffic flow and revenue under the conventional business model from 2023 to 2036.

In predicting the annual traffic flow, we also take into account the highway's maximum capacity, set at 4,000 vehicles per hour for Guanghe Highway's dual carriageway. This implies a maximum annual traffic flow of 35.04 million vehicles. In cases where annual traffic surpasses this maximum, we use 35 million vehicles. We then employ Matlab's curve fitting tool for polynomial fitting to establish the relationship between annual traffic flow and income.

Based on this, we can estimate the corresponding revenue based on the annual traffic flow of the new smart highway under the proposed business model. Prior to this, we need to estimate the traffic flow under the new model. We primarily consider the impact of road traffic efficiency and penetration rate on the annual traffic flow. As per Figure \ref{fig6} simulation results, with a GV penetration rate below 50\%, average vehicle velocity increases by roughly 3\%. Beyond 50\% penetration, the average vehicle speed surges by approximately 30\%. We base GV penetration rate growth on the new energy vehicle penetration rate in China, which reached 27.6\% in 2022, a 12.6 percentage point increase from 2021. We assume a 10-percentage-point annual GV penetration rate increase.

\begin{figure*}[t]
\begin{center}
\includegraphics[width=17cm]{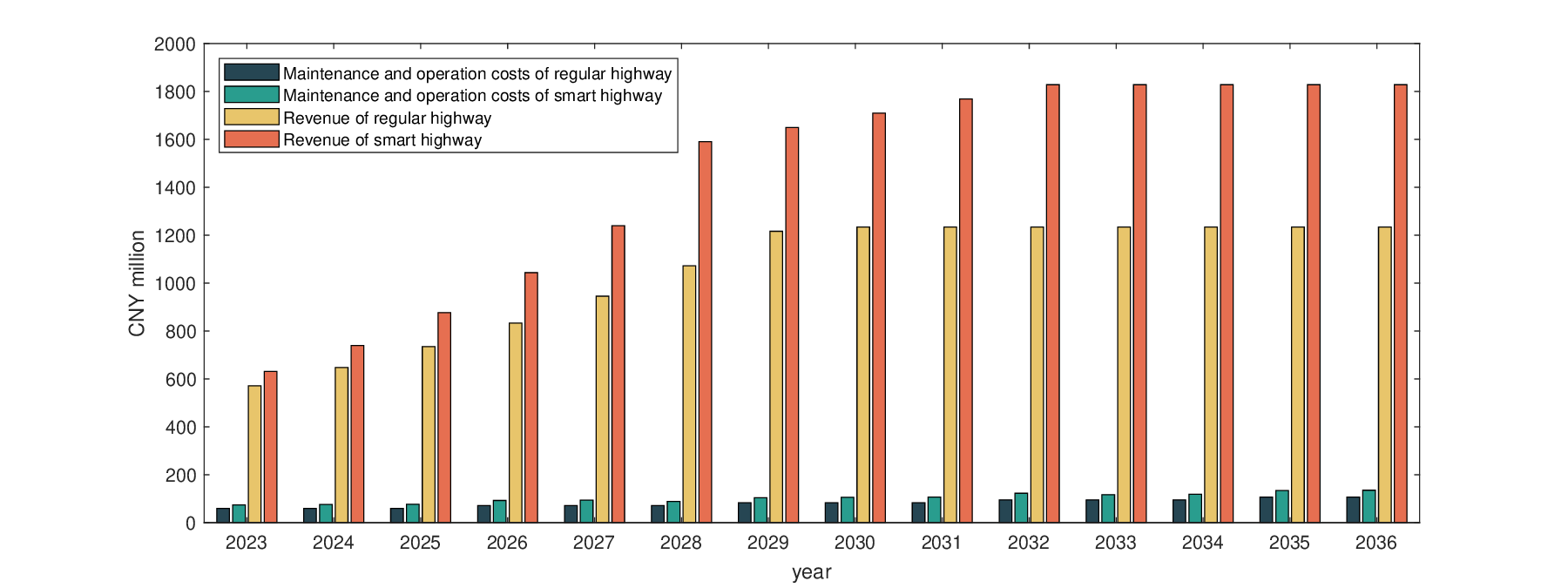}
\end{center}
\caption{The predicted maintenance and operation costs and the revenue, taking Guanghe Expressway Guangzhou section as an example, for both regular highway and smart highway in the future years.}%
\label{fig7}
\end{figure*}

Furthermore, we assume that the annual traffic flow growth rate of the smart highway is equal to the percentage increase in average vehicle velocity corresponding to the GV penetration rate. For instance, in the first year with a 10\% GV penetration rate, the smart highway's annual traffic flow compared to the regular highway increases by 3\%. As GV penetration rises, in the sixth year at a 60\% penetration rate, the annual traffic flow increases by 30\%. Lastly, smart highway revenue includes guided fees, set at CNY 0.2 per kilometer. Considering the issue of purchasing power, we assume that the number of GV vehicles purchasing guidance services accounts for 70\% of the total GV count. As shown in Figure \ref{fig7}, the diagram illustrates the annual costs and revenues from 2023 to 2036 for both the regular business model and the novel business model.

\paragraph{Cost-Benefit Ratio}
The benefit-cost ratio (BCR) is an indicator used to evaluate the economic benefits of a project, comparing the expected benefits of the project to the investment cost \cite{shively2013overview}. Generally, a higher benefit-cost ratio indicates that the expected benefits of the project are more favorable relative to the investment cost.

\begin{table}[H]
\large
\caption{\textbf{BCR of different highway types}}
\centering
\renewcommand{\arraystretch}{1.5}
\scalebox{0.75}{
\begin{tabular}{cccc}
\toprule
Model & Cost & Net Benefit & BCR\\
\midrule
Regular highway & 114224.8048 & 1351606.198 & 11.83286065 \\
Smart highway  & 147468.0566 & 1894420.077 & 13.09588066 \\
\bottomrule
\end{tabular}
}
\label{t1}
\end{table}

We calculate the benefit-cost ratios for two different business models: regular highway and smart highway with guided fees. The calculation results are shown in Table \ref{t1}. In the table, ``cost'' refers to the sum of total costs from 2023 to 2036. Particularly, we have taken into account the additional costs of deploying and installing roadside perception and communication devices in the smart highway, which amounts to an extra CNY 28.1034888 million. On the other hand, ``net benefit'' represent the sum of annual revenues minus costs from 2023 to 2036. The final results demonstrate that smart highway with guided fees exhibits a higher benefit-cost ratio compared to regular highway. This implies that adopting our proposed new business model can yield greater economic benefits.

\section{Conclusions}

The rise of vehicle-road cooperation tech in autonomous driving research, with strengths in perception and decision-making, faces practical challenges. Its high cost hinders widespread adoption, and its actual benefits remain unclear.  To tackle these challenges, this paper has proposed a profitable and novel business model that utilizes smart-road vehicle-guiding architecture for autonomous driving.  This innovative business model has brought about a win-win case for all stakeholders involved. The experiments have demonstrated that GV has exhibited a higher level of safety compared to AV.  As the penetration rate of GV increases, there has been an observable increase in average vehicle speed, indicating that the widespread adoption of GV contributes to enhanced traffic efficiency. Using the CBA methodology, we conclude that despite higher initial costs, vehicle-road cooperation offers a superior benefit-cost ratio, showcasing its economic advantages. Overall, this innovative business model advances the commercialization of autonomous driving technology, bringing benefits to the transportation industry and society as a whole.

\section*{Acknowledgment}
The authors would like to thank the editor and the anonymous reviewers for their careful reading and valuable suggestions that helped to improve the quality of this manuscript.
\bibliographystyle{IEEEtran}
\bibliography{IEEEabrv,mds}
\begin{IEEEbiography}[{\includegraphics[width=1in,height=1.25in,clip,keepaspectratio]{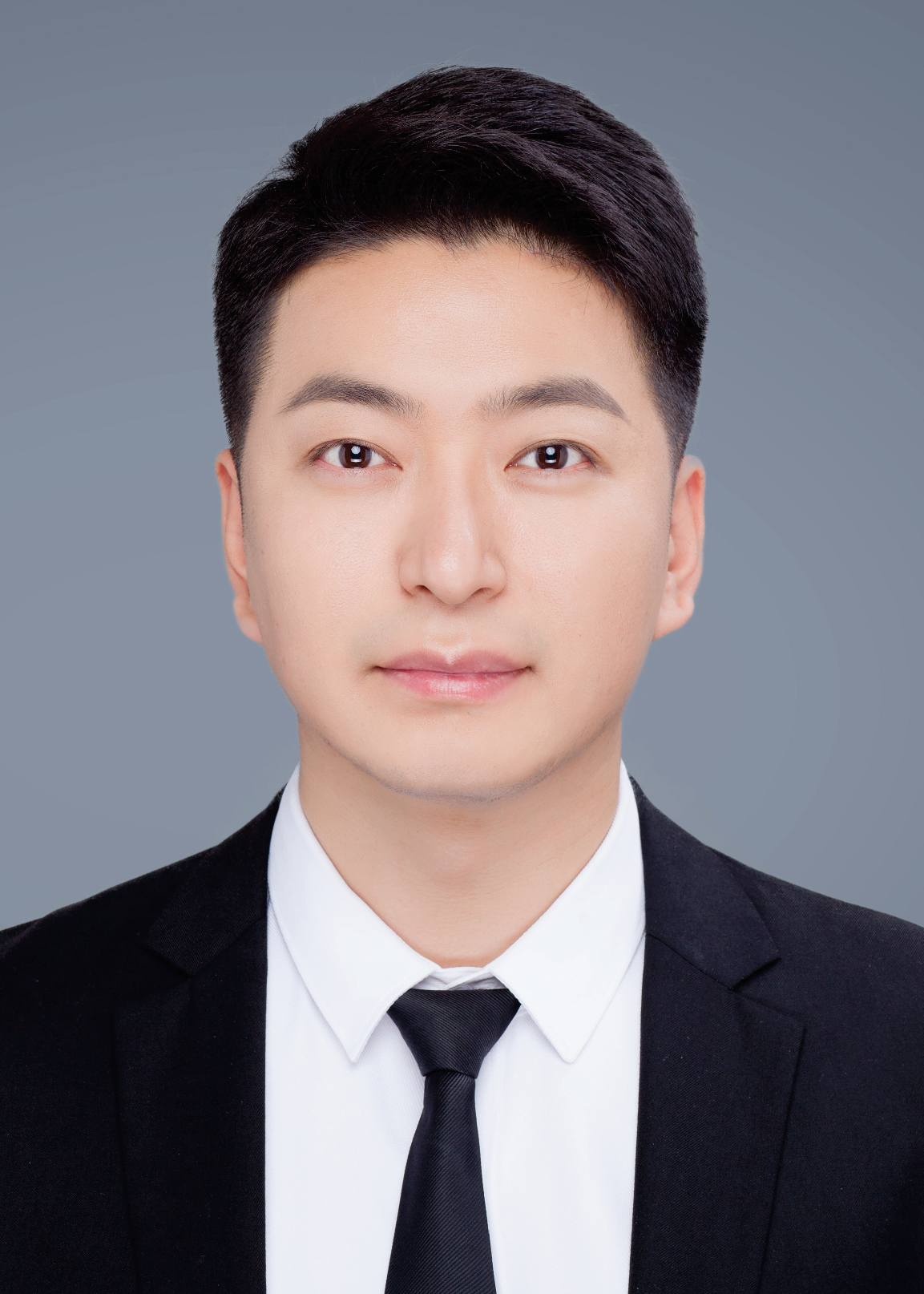}}]{Rui Chen}
(Member, IEEE) received the B.S., M.S., and Ph.D. degrees in communications and information systems from Xidian University, Xi'an, China, in 2005, 2007, and 2011, respectively. From 2014 to 2015, he was a Visiting Scholar with Columbia University, New York. He is currently a Professor and Ph.D. Supervisor with the School of Telecommunications Engineering, Xidian University. His research interests include broadband wireless communication systems, array signal processing, and intelligent transportation systems.

\end{IEEEbiography}
\begin{IEEEbiography}[{\includegraphics[width=1in,height=1.25in,clip,keepaspectratio]{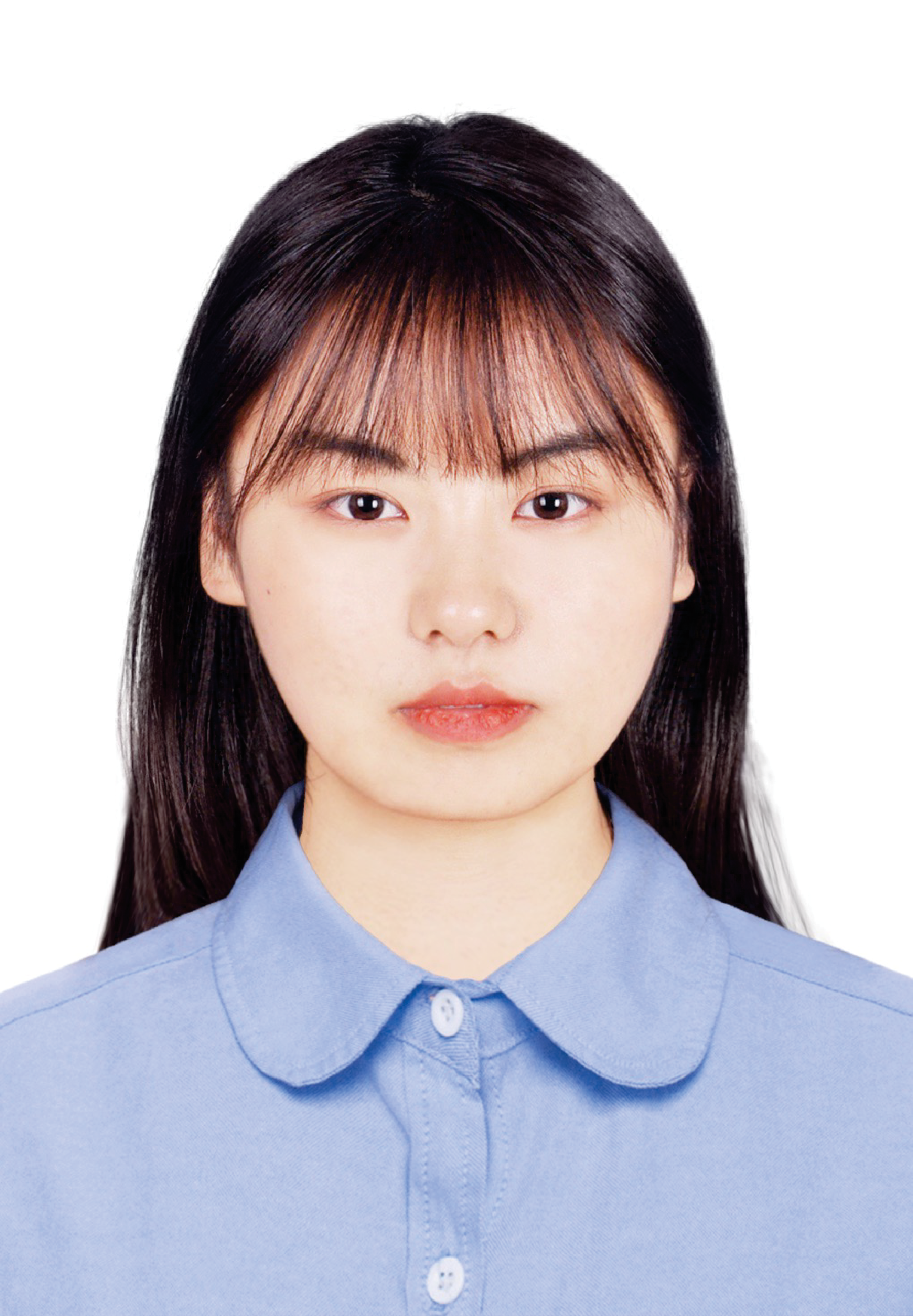}}]{Lu Gao}
(Graduate Student Member, IEEE) received the B.S. degree in computer science and technology from the North China University of Science and Technology, Hebei, China, in 2022.  She is currently pursuing the M.S. degree in communications and information systems from Xidian University, Xi'an, China. Her research interests include radar data processing and application.
\end{IEEEbiography}
\begin{IEEEbiography}[{\includegraphics[width=1in,height=1.25in,clip,keepaspectratio]{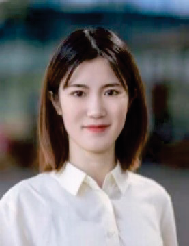}}]{Yutian Liu}(Graduate Student Member, IEEE) obtained the B.S. degree in Engineering Mechanic from Harbin Institute of Technology and the M.S. degree in Aeronautical and Astronautical Science and Technology from Harbin Engineering University. She is currently a PhD candidate in the group of Urban Planning and Transportation from Eindhoven University of Technology (TU/e). Her research interests include intelligent transportation system, traffic flow prediction based on graph neural networks (GCN) and robustness on deep learning models.
\end{IEEEbiography}

\begin{IEEEbiography}[{\includegraphics[width=1in,height=1.25in,clip,keepaspectratio]{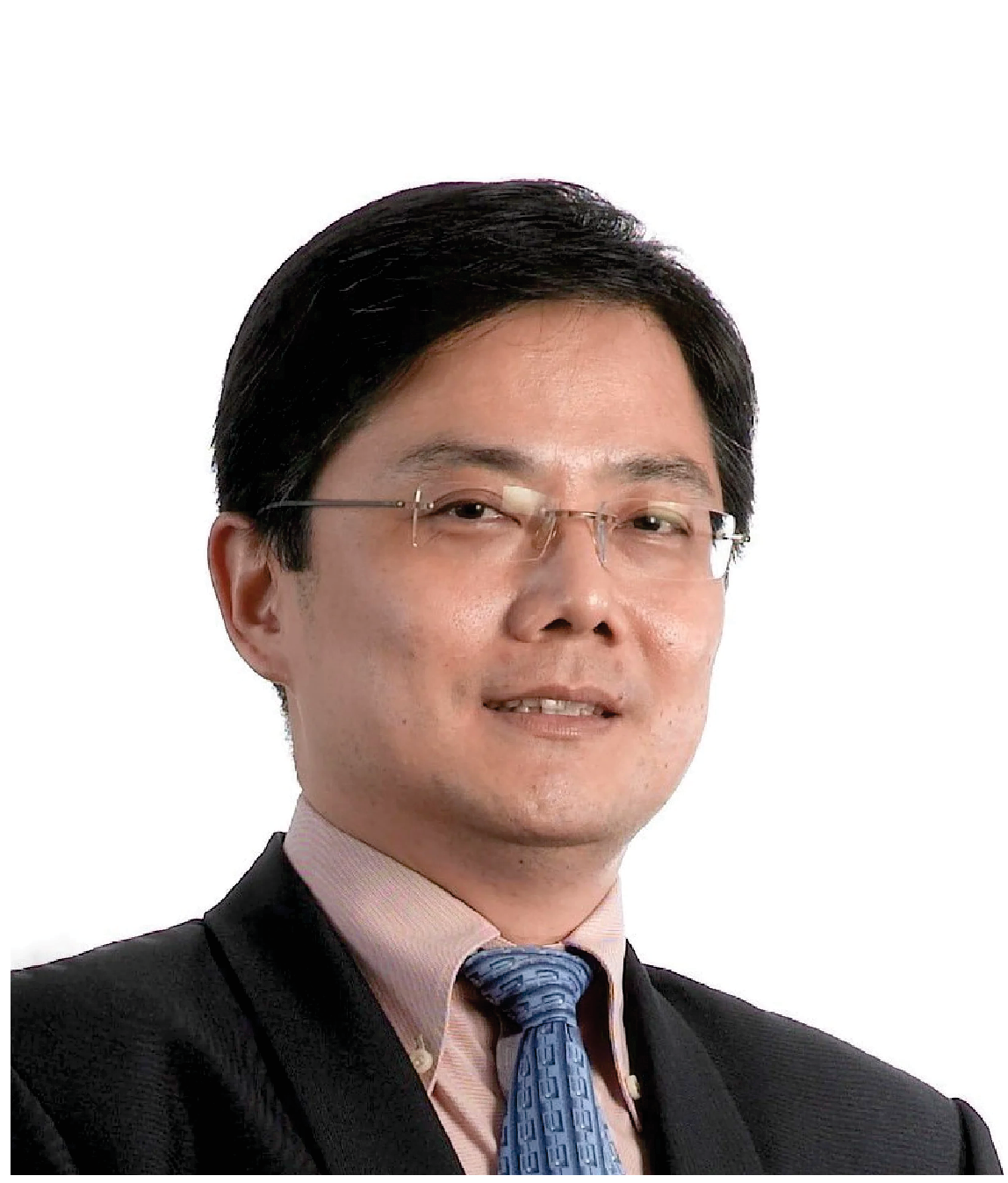}}]{Yong Liang Guan}
(Senior Member, IEEE) obtained his PhD degree from the Imperial College London, UK, and Bachelor of Engineering degree with first class honours from the National University of Singapore. He is now a Professor of Communication Engineering at the School of Electrical and Electronic Engineering, Nanyang Technological University (NTU), Singapore, where he leads the Continental-NTU Corporate Research Lab and led the successful deployment of the campus-wide NTU-NXP V2X Test Bed. His research interests broadly include coding and signal processing for communication systems and data storage systems.
\end{IEEEbiography}

\begin{IEEEbiography}[{\includegraphics[width=1in,height=1.25in,clip,keepaspectratio]{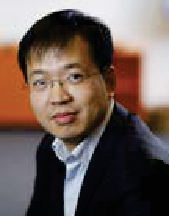}}]{Yan Zhang}
(Fellow, IEEE) received the Ph.D. degree in electrical and electronic engineering from the School of Electrical and Electronics Engineering, Nanyang Technological University, Singapore, in 2005. He is currently a Full Professor with the Department of Informatics, University of Oslo, Oslo, Norway. His research interests include next-generation wireless networks leading to 5G beyond/6G, green, and secure cyber-physical systems (e.g., smart grid and transport).
\end{IEEEbiography}
\end{document}